\documentstyle[twocolumn,epsf]{jpsj}
\def\runtitle{ 
Oscillator Strength of Metallic Carbon Nanotubes 
}
\def\runauthor{Hideo {\sc Yoshioka}}
\def\d{{\rm d}}
\def\lan{\left\langle}
\def\ran{\right\rangle}

\def\e{{\rm e}}

\def\ggs{\buildrel\textstyle > \over {\hbox{\raise0.2ex\hbox{$\sim$}}}}
\def\lls{\buildrel\textstyle < \over {\hbox{\raise0.2ex\hbox{$\sim$}}}}
\def\gsim{\,\lower0.75ex\hbox{$\ggs$}\,}
\def\lsim{\,\lower0.75ex\hbox{$\lls$}\,}

\def\im{{\rm i}}
\def\ie{{\it i.e.}, }

\def\al{\alpha}
\def\eg{{\it e.g.}}
\def\jo #1#2#3#4{#1 {\bf #2} (#3) #4}   

\def\PRB{Phys.\ Rev.\ B}
\def\PRL{Phys.\ Rev.\ Lett.}

\def\APL{Appl.\ Phys.\ Lett.}

\def\JPC{J.\ Phys.\ C}

\def\JPSJ{J.\ Phys.\ Soc.\ Jpn.}

\def\NAT{Nature (London)}
\def\EPJB{Eur.\ Phys.\ J.\ B}

\title
{
Oscillator Strength of Metallic Carbon Nanotubes 
}

\author{
Hideo {\sc Yoshioka}\hspace{-0.5mm}\footnote{E-mail: yoshioka@phys.nara-wu.ac.jp}\hspace{0.5mm}
}

\inst{
Department of Physics, Nara Women's University, Nara 630-8506
}

\recdate{June 30, 2000}

\abst{
Based on the tight binding method with hopping integral 
between the nearest-neighbor atoms, 
an oscillator strength $\int_0^{\infty} \d \omega {\rm Re} \sigma (\omega)$
is discussed for armchair and metallic zigzag carbon nanotubes.   
The formulae of the oscillator strength 
are derived for both types of nanotubes 
and are compared with the result obtained by 
a linear chain model. 
In addition, the doping dependence is   
investigated in the absence of Coulomb interaction.
It is shown that 
the oscillator strength of each carbon nanotube
shows qualitatively the same doping dependence, but 
the fine structure is different
due to it's own peculiar band structure.   
Some relations independent of the radius of the tube are derived, 
and a useful formula for determining the amount of doping is proposed. 
}

\kword{carbon nanotubes, oscillator strength, doping
}

\begin{document}
\sloppy
\maketitle



A carbon nanotube (CN) 
is composed of a coaxially rolled 
graphite sheet.\cite{Iijima}
The materials are characterized by two integers, $(N_+, N_-)$, 
corresponding to a wrapping vector along the waist, 
$\vec w = N_+ \vec a_+ + N_- \vec a_-$, 
where $\vec a_{+}$ and $\vec a_{-}$ are primitive lattice vectors of 
the graphite and $|\vec a_{\pm}| = a$.
It has been shown that the CN's have peculiar 
band structures.\cite{Hamada,Saito-I,Saito-II} 
When $N_+ - N_- = 0$ mod 3, 
the metallic one-dimensional (1D) 
dispersions appear near the center of the bands. 
The low energy properties less than $v_0 / R$ 
($v_0$ : Fermi velocity, $R$ : radius of the tube ) is described 
by taking account of only the metallic 1D
dispersions.\cite{EG-I,Kane,EG-II,YO,OY}
Correlation effects obtained by such a treatment have been observed
in the transport experiment.\cite{Bockrath}    

Carrier doping to CN's has been done by doping the electron-donor 
(\eg, potassium, rubidium) or electron-acceptor (\eg, iodine, bromine).
\cite{Lee1,Rao,Lee2,Claye} 
It has been experimentally observed 
that the doping changes the properties of CN's. 
In bundles of single wall CN's, 
bromine and potassium doping decrease the resistivity 
at 300K up to a factor of 30, and enlarge 
the region where the temperature coefficient of resistance
is positive.\cite{Lee1}
The similar behavior is observed in a potassium-doped
single rope.\cite{Lee2}   
Change of the Ramann spectra has been observed in  
the bundles of CN's with doping of 
K, Rb and Br$_2$.\cite{Rao}  
Enhancement of spin susceptibility due to potassium-doping
has been also reported.\cite{Claye}  
As an another method for doping, a downward shift of the Fermi level due to 
the gold substrate has been reported 
by scanning tunneling spectroscopy.\cite{Wildoer}  
 
The oscillator strength  
$\int_0^{\infty} \d \omega {\rm Re} \sigma (\omega)$, 
where $\sigma (\omega)$ is the optical conductivity, 
is closely related to the amount of carrier. 
Irrespective of the presence of the mutual interaction,
as far as the kinetic energy is expressed by  the quadratic dispersion,   
the oscillator strength is given as follows, 
\begin{equation}
\int_0^{\infty} \d \omega {\rm Re} \sigma (\omega) 
= \frac{\pi}{2} \frac{n e^2}{m} , 
\label{eqn:free-sum}  
\end{equation}  
where $n$ and $m$ are the electron density and the mass, respectively.
On the other hand, the oscillator strength 
is different from eq.(\ref{eqn:free-sum}) 
in the case of tight binding models.\cite{sum}
As the simplest system, we consider a linear chain 
model with the nearest neighbor hopping, $t_{i,i+1}$,
whose kinetic energy is given by 
$T = - \sum_{i,s} \{ t_{i,i+1} c^\dagger_{i,s} c_{i+1,s} + h.c. \}$. 
The oscillator strength is calculated as,\cite{sum}  
\begin{equation}
\int_0^{\infty} \d \omega {\rm Re} \sigma (\omega) 
= - \frac{\pi}{2} \frac{e^2 d}{N_L} \lan T \ran, 
\label{eqn:HM-sum}  
\end{equation}  
where $N_L$ is the number of the lattice sites, $d$ is a lattice spacing and 
$\lan \cdots \ran$ is the thermal average.
The quantity, $\lan T \ran$, is not proportional to the electron density. 
For example, in case of $t_{i,i+1} = t$ without the mutual interaction,
eq.(\ref{eqn:HM-sum}) reduces to 
$2 e^2 t d \sin (\pi d n / 2)$ at the absolute zero temperature.  
The difference between 
eqs.(\ref{eqn:free-sum}) and (\ref{eqn:HM-sum}) with respect to 
carrier dependence 
results from distinction of their band structures. 
  
It is obvious that the quantities for the high energy scale 
such as the oscillator strength cannot be described by 
the theory in which only the metallic dispersions 
are taken into account. 
Ando has been calculated the optical conductivity 
of CN's by using the effective mass theory, 
which is valid for the energy scale less than 
$v_0/a$.\cite{Ando}
He predicted that the absorption edge is shifted to the higher 
energy side due to Coulomb interaction, which has been 
observed in the recent experiment.\cite{Ichida} 
Thus the effective mass theory succeeds in describing 
the low energy physics very well. 
However, it is questionable whether the theory is effective for 
discussing the oscillator strength and it's dependence 
on the amount of the carrier.   

In the present paper,
using the tight binding method, 
we derive the formulae of the oscillator strength 
of $(N,N)$ armchair CN's 
and metallic $(N,0)$ zigzag CN's. 
The formulae are compared with the result 
of the linear chain model, eq.(\ref{eqn:HM-sum}).  
In addition, 
the doping dependence of the 
oscillator strength is discussed in detail\cite{Ando-II}  
in the absence of Coulomb interaction.
We take a unit of $\hbar = c = 1$.

\begin{figure}
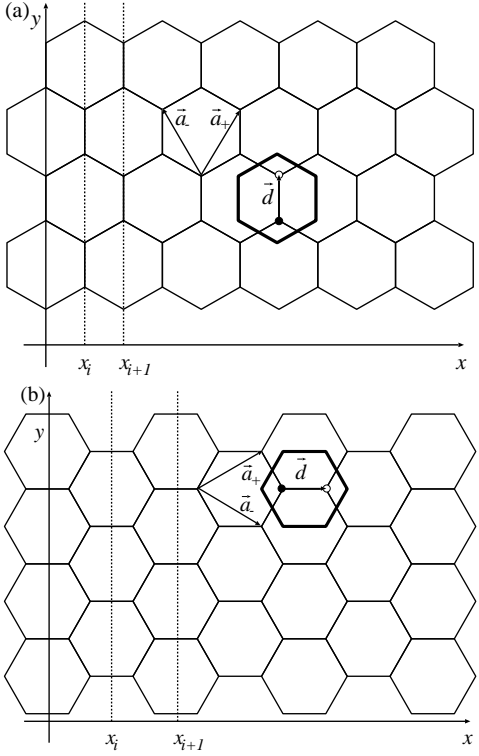

\vspace{2em}
\centerline{\epsfxsize=6.3cm\epsfbox{AN.eps}}
\centerline{\epsfxsize=6.3cm\epsfbox{ZN.eps}}
\caption{ Carbon atoms in armchair nanotubes (a)
and zigzag nanotubes (b)
where the $x$ axis points along the tube. 
Here $\vec a_{\pm}$ are two primitive lattice vectors of a graphite, 
and $|\vec a_{\pm}| = a$.  
The hexagon shown by the thick line is the unit cell and 
the black (white) circle denotes the sublattice $p=+(-)$.
} 
    \label{fig:CN}
    \end{figure}

We consider the armchair and zigzag CN's shown in
Fig.\ref{fig:CN} (a) and (b), respectively.
Here the directions of the tube and of the waist are denoted as $x$ and
$y$, respectively. 
An electric field is applied to the $x$-direction. 
The kinetic energy of the armchair CN, ${\cal H}_{\rm k}^{AN}$, 
in the presence of the time dependent vector potential 
along the $x$-direction $A(x,u)$,  
is given as, 
\begin{eqnarray}
& & {\cal H}_{\rm k}^{AN} = \sum_{{\vec r}_i, s} 
\left\{ -t a^\dagger_{-,s}({\vec r}_i) a_{+,s}({\vec r}_i) + h.c. \right\} \nonumber \\
&+& \sum_{{\vec r}_i, s} 
\left\{ -t \e^{\im e A(x_i,u) a / 2} 
a^\dagger_{+,s}({\vec r}_i) a_{-,s}({\vec r}_i - {\vec a}_+) + h.c. \right\} \nonumber \\
 &+& \sum_{{\vec r}_i, s} 
\left\{ -t \e^{\im e A(x_i,u) a / 2} 
 a^\dagger_{-,s}({\vec r}_i) a_{+,s}({\vec r}_i + {\vec a}_-) + h.c. \right\}, 
\label{eqn:Hk}
\end{eqnarray}
where $t$ denotes the hopping integral between the nearest-neighbor atoms, 
and $a^\dagger_{p,s} ({\vec r}_i)$
is the creation operator of the electron with spin $s$ 
at the location, ${\vec r}_i - p {\vec d}/2$ ($p=\pm$).
The electric field is given by $- \partial_u A(x,u)$.

The current operator $I^{AN}(x_i)$  
is obtained by differentiating eq.(\ref{eqn:Hk}) 
in terms of $A(x_i,u)$.  
Up to the first order of $A(x_i,u)$, 
\begin{eqnarray}
I^{AN} (x_i)
&=& \im e t \sum_{y_i,s} 
\Big\{
a^\dagger_{+,s}(\vec r_i) a_{-,s}(\vec r_i - \vec a_+) 
\nonumber \\
&+& a^\dagger_{-,s}(\vec r_i) a_{+,s}(\vec r_i + \vec a_-)
- h.c. 
\Big\} \nonumber \\
&-& \frac{e^2 t A(x_i,u) a}{2} \sum_{y_i,s} 
\Big\{
a^\dagger_{+,s}(\vec r_i) a_{-,s}(\vec r_i - \vec a_+) 
\nonumber \\
&+&  a^\dagger_{-,s}(\vec r_i) a_{+,s}(\vec r_i + \vec a_-)
+ h.c. 
\Big\} . 
\label{eqn:C-AN}
\end{eqnarray}
On the other hand, the kinetic energy of the zigzag CN, 
\begin{eqnarray}
& & {\cal H}_{\rm k}^{ZN} = \sum_{{\vec r}_i, s} 
\left\{ -t \e^{\im e A(x_i,u) a / \sqrt{3} } 
a^\dagger_{-,s}({\vec r}_i) a_{+,s}({\vec r}_i) + h.c. \right\} \nonumber \\
&+& \sum_{{\vec r}_i, s} 
\left\{ -t \e^{\im e A(x_i,u) a / \sqrt{12} } 
a^\dagger_{+,s}({\vec r}_i) a_{-,s}({\vec r}_i - {\vec a}_+) + h.c. \right\} \nonumber \\
 &+& \sum_{{\vec r}_i, s} 
\left\{ -t \e^{\im e A(x_i,u) a / \sqrt{12} } 
 a^\dagger_{+,s}({\vec r}_i) a_{-,s}({\vec r}_i - {\vec a}_-) + h.c. \right\}, 
\label{eqn:Hk-ZN}
\end{eqnarray}
leads to the current operator, 
\begin{eqnarray}
& & I^{ZN}(x_i) = 
\frac{\im e t}{3} \sum_{y_i,s} 
\Big\{
2 a^\dagger_{-,s}(\vec r_i) a_{+,s}(\vec r_i) 
\nonumber \\
&+& a^\dagger_{+,s}(\vec r_i) a_{-,s}(\vec r_i - \vec a_+)
+ a^\dagger_{+,s}(\vec r_i) a_{-,s}(\vec r_i - \vec a_-)
- h.c. 
\Big\} \nonumber \\
&-& \frac{e^2 t A(x_i,u) a}{6 \sqrt{3}}  \sum_{y_i,s} 
\Big\{
4 a^\dagger_{-,s}(\vec r_i) a_{+,s}(\vec r_i) 
+ a^\dagger_{+,s}(\vec r_i) a_{-,s}(\vec r_i - \vec a_+) \nonumber \\
&+& a^\dagger_{+,s}(\vec r_i) a_{-,s}(\vec r_i - \vec a_-)
+ h.c.
\Big\} . 
\label{eqn:C-ZN}
\end{eqnarray}
The current density operator per unit length along the waist,  
${j}_{2D} (x_i)$, and that per unit area, 
${j}_{3D} (x_i)$, 
are given by 
${j}^{C}_{2D} (x_i) = {I}^{C}(x_i)/(2 \pi R)$
and ${j}^{C}_{3D} (x_i) = {I}^{C}(x_i)/(\pi R^2)$, 
where $R = \sqrt{3} N a / (2 \pi)$ and $R = N a / (2 \pi)$ for 
$C$ = $AN$ (armchair CN's) and  $C = ZN$  (zigzag CN's),  respectively.

In order to compare the oscillator strength between the linear chain
model
and the CN,  
we calculate the oscillator strength corresponding to the 1D currents,
eqs.(\ref{eqn:C-AN}) and (\ref{eqn:C-ZN}).
The diamagnetic component of the current, which is proportional to 
$A(x,u)$, 
leads to the oscillator strength $Z^C_{1D}$ : 
\begin{eqnarray}
Z^{C}_{1D} &=& -\frac{\pi}{2} \lan X^{C} \ran, 
\end{eqnarray}  
where $\lan \cdots \ran$ expresses the thermal average 
in the absence of the vector potential, and $X^C$ ($C = AN$, $ZN$) are
defined as, 
\begin{eqnarray}
X^{AN} &=& - \frac{e^2 t}{L} \left(\frac{a^2}{4}\right)
\sum_{\vec r_i s}
\Big\{
a^\dagger_{+,s}(\vec r_i) a_{-,s}(\vec r_i - \vec a_+) \nonumber \\
&&\hspace{2em} +
a^\dagger_{-,s}(\vec r_i) a_{+,s}(\vec r_i + \vec a_-) + h.c. 
\Big\} , 
\label{eqn:X-AN} 
\end{eqnarray}
and
\begin{eqnarray}
& & X^{ZN} = - \frac{e^2 t}{L} \left(\frac{a^2}{12}\right)
\sum_{\vec r_i s}
\Big\{
4 a^\dagger_{-,s}(\vec r_i) a_{+,s}(\vec r_i ) \nonumber \\
&+& a^\dagger_{+,s}(\vec r_i) a_{-,s}(\vec r_i - \vec a_+) 
+ a^\dagger_{+,s}(\vec r_i) a_{+,s}(\vec r_i - \vec a_-) + h.c. 
\Big\} , \nonumber \\
\label{eqn:X-ZN} 
\end{eqnarray}
with $L$ being the length of the tube.  
The oscillator strength 
 corresponding to the 2D and 3D current densities 
are given by $Z_{2D}^{C} = Z_{1D}^{C}/(2\pi R)$ 
and $Z_{3D}^{C} = Z_{1D}^{C}(\omega)/(\pi R^2)$. 
The quantities $X^{AN}$  and $X^{ZN}$ are, respectively,  
not proportional to 
the kinetic energies eq.(\ref{eqn:Hk}) and eq.(\ref{eqn:Hk-ZN})
in the case of $A(x,u) = 0$.  
It is due to the fact that 
the CN is not the linear chain system. 

Now we calculate $Z^{C}_{1D}$ as a function of hole doping, 
$\delta \equiv 1 - N_e/N_c$, where $N_e$ ($N_c$) is the number of electrons
(carbon atoms),   
in the absence of Coulomb interaction. 
In this case, 
the intensity of the optical conductivity 
exists within the band width, \ie $0 \leq \omega \leq 6 t$.  
The quantities, $\lan X^{C} \ran$ and $\delta$, 
are determined by the following equations
as a function of the chemical potential, $\mu$.   
For armchair CN's, 
\begin{eqnarray}
& & \lan X^{AN} \ran = \frac{t e^2 a^2}{4 \pi} 
\int_{- 2 \pi/a}^{2 \pi / a} \d K_x \nonumber \\  
&\times& \sum_{K_y} \sum_{\al=\pm} 
\al Z^{AN} (\vec K) f( \al \xi^{AN}(\vec K) - \mu) , \\ 
& & \delta = \frac{a}{4 N \pi} 
\int_{- 2 \pi/a}^{2 \pi / a} \d K_x \nonumber \\ 
&\times& \sum_{K_y} \sum_{\al = \pm}
\Big\{ 
f( \al \xi^{AN}(\vec K)) - f(\al \xi^{AN}(\vec K)- \mu) 
\Big\} ,    
\end{eqnarray}
with
\begin{eqnarray}
Z^{AN}(\vec K) &=& 2 t 
\cos (K_x a/2) \nonumber \\
&& \hspace{-5em} \times \left\{ 2 \cos ({K_x a}/{2}) + 
\cos ({K_y \sqrt{3} a}/{2}) \right\} / \xi^{AN} (\vec K),    \\
\xi^{AN} (\vec K) &=& t
\Big\{
1 + 4 \cos^2 ({K_x a}/{2}) \nonumber \\
& & + 4 \cos ({K_x a}/{2}) 
\cos ({K_y \sqrt{3} a}/{2}) 
\Big\}^{1/2} , 
\end{eqnarray}
where ${\vec K} = (K_x, K_y)$, $K_y = 2 \pi n / (\sqrt{3} N a)$ 
with $n$ being $- [N /2], \cdots, [N/2]$ for odd $N$ 
or $- N /2, \cdots, N/2 - 1$ for even $N$ and
$f(\epsilon)$ is the Fermi function.
On the other hand, for zigzag nanotubes, 
\begin{eqnarray}
& &\lan X^{ZN} \ran = \frac{t e^2 a^2}{12 \pi} 
\int_{0}^{4 \pi / (\sqrt{3}a)} \d K_x \nonumber \\  
&\times& \sum_{K_y} \sum_{\al=\pm} 
\al Z^{ZN} (\vec K) f( \al \xi^{ZN}(\vec K) - \mu) , \\ 
\delta &=& \frac{\sqrt{3}a}{4 N \pi} 
\int_{0}^{4 \pi / \sqrt{3} a} \d K_x \nonumber \\ 
&\times& \sum_{K_y} \sum_{\al = \pm}
\Big\{ 
f( \al \xi^{ZN}(\vec K)) - f(\al \xi^{ZN}(\vec K)- \mu) 
\Big\} ,    
\end{eqnarray}
with
\begin{eqnarray}
Z^{ZN}(\vec K) &=& 2 t
\Big\{ 2 + 5 \cos (K_x \sqrt{3} a/2) \cos K_y a /2 \nonumber \\
 &+& 2 \cos^2 K_y a / 2 \Big\} 
/ \xi^{ZN} (\vec K),    \\
\xi^{ZN} (\vec K) &=& t
\Big\{
1 + 4 \cos^2 ({K_y a}/{2}) \nonumber \\
& & + 4 \cos ({K_y a}/{2}) 
\cos ({K_x \sqrt{3} a}/{2}) 
\Big\}^{1/2} , 
\end{eqnarray}
where $K_y = 2 \pi n / (N a)$ with $n$ being 
$- [N /2], \cdots, [N/2]$ for odd $N$  or 
$- N /2, \cdots, N/2 - 1$ for even $N$. 
Note that $\pm \xi^{AN}(\vec K)$  ($\pm \xi^{ZN}(\vec K)$)
are the energy dispersions of the armchair (zigzag) CN, 
and those with $K_y = 0$ ($K_y = \pm 2 \pi/3a$) can be zero.
Hereafter, such dispersions are called as center metallic bands.  

\begin{figure}
\vspace{2em}
\centerline{\epsfxsize=6.3cm\epsfbox{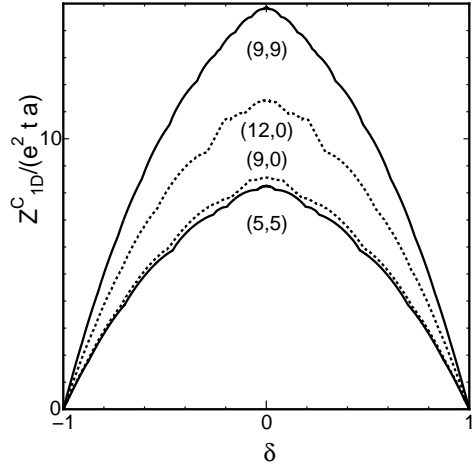}}
    \caption{$Z_{1D}^{C}$ in unit of $e^2 t a$ of $(N,N)$ armchair and $(N,0)$ zigzag CN's
as a function of $\delta$ for the several choices
   of $N$.}
    \label{fig:z1d}
    \end{figure}

$Z_{1D}^{C}/(t e^2 a)$ is shown in Fig.\ref{fig:z1d} 
as a function of $\delta$
at the absolute zero temperature. 
  \begin{figure}
\centerline{\epsfxsize=6.3cm\epsfbox{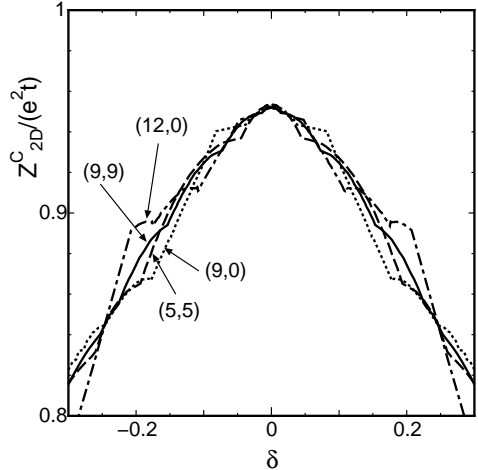}}
    \caption{$Z_{2D}^{C}$ in unit of $e^2 t$ of $(N,N)$ armchair and $(N,0)$ zigzag CN's
as a function of $\delta$ for the several choices
   of $N$.}
    \label{fig:z2d}
    \end{figure}
$Z^C_{1D}$ has a maximum at half-filling ($\delta = 0$) :   
It decreases with increasing
 $|\delta|$ and vanishes 
when the bands are empty or full ($\delta = 1$ or $-1$). 
The fine structures are due to the Van Hove singularities
of the density of states.  
We find that,  
within  numerical accuracy, 
$Z^C_{2D}$ in the absence of doping ($\delta = 0$) 
is independent of $N$ and has the same value for the armchair and 
zigzag CN's, \ie  
$Z^{AN}_{2D} = Z^{ZN}_{2D} \simeq 0.952 e^2 t$
as is shown in Fig.\ref{fig:z2d}. 
It seems to be related to the fact that 
the number of the states less than Fermi energy at $\delta = 0$
for the system with the fixed tube length 
is proportional to $R$ and independent of chirality.  
The doping dependence, however, differs for each nanotube 
due to it's own peculiar band structure. 

Next, let us consider the doping dependence, 
$\Delta Z^C_{1D} \equiv Z^C_{1D}|_{\delta = 0} - Z^C_{1D}$, 
in detail. 
In the case where the chemical potential stays 
at only the center metallic bands,
the quantities, $\delta$ and $\Delta Z^C_{1D}$, can be calculated as
follows : 
For armchair CN's, 
\begin{eqnarray}
\delta &=& \frac{K_+ - K_-}{2 N \pi} , 
\label{eqn:DAN}\\
\Delta Z^{AN}_{1D} &=& t e^2 a 
\left\{
\sqrt{3} - \sin K_+ a /2 - \sin K_- a /2 
\right\} , 
\label{eqn:DZAN}
\end{eqnarray}
where $\cos K_\pm a /2 = (\pm \mu/t -1)/2$.
And for zigzag nanotubes, 
\begin{eqnarray}
\delta &=& \frac{\sqrt{3}a}{2 N \pi} ( K_+ - K_- ), 
\label{eqn:DZN}\\
\Delta Z^{ZN}_{1D} &=& \frac{5}{3\sqrt{3}}t e^2 a 
\left\{
2 - \sin \frac{K_+ \sqrt{3}a}{4} 
- \sin \frac{K_- \sqrt{3}a}{4} 
\right\} , \nonumber \\
\label{eqn:DZZN} 
\end{eqnarray}
where $\cos K_\pm \sqrt{3} a /4 = \pm \mu/(2t)$.
In the case of $|\mu / t| \ll 1$, 
eqs.(\ref{eqn:DAN})-(\ref{eqn:DZZN}) lead to 
\begin{eqnarray}
\Delta Z^{AN}_{1D} &\simeq& 
  t e^2 a
\frac{N^2 \pi^2}{2 \sqrt{3}} \delta^2 , 
\label{eqn:DZANAP}\\
\Delta Z^{ZN}_{1D} &\simeq& 
  t e^2 a
\frac{5 N^2 \pi^2}{48 \sqrt{3}} \delta^2 . 
\label{eqn:DZZNAP}
\end{eqnarray}
Thus, $\Delta Z^C_{1D}$ is proportional to $\delta^2$ and 
the square of the radius of the tube. 
Then $\Delta Z^C_{3D} \equiv \Delta Z^C_{1D}/(\pi R^2)$
is independent of the radius as 
$\Delta Z^{AN}_{3D} = (t e^2 / a) (2 \pi^3/3 \sqrt{3}) \delta^2$ and 
$\Delta Z^{ZN}_{3D} = (t e^2 / a) (5 \pi^3/12 \sqrt{3}) \delta^2$.
The values of both equations are close to each other.   
The quantity obtained experimentally 
is nothing but $Z_{3D}$ 
because the reflectance observed experimentally
is related to the optical conductivity corresponding to 
the 3D current density.   
Therefore, from the above equations, 
the amount of doping may be well determined 
even when CN's with different chirality and radius coexist
if the CN's with the another chirality have values close to 
$\Delta Z^{AN}_{3D}$ and/or $\Delta Z^{ZN}_{3D}$. 

In summary, we have derived the formulae of the oscillator strength 
of the armchair and metallic zigzag CN's, and compared with 
the result of the linear chain model.   
Dependence of the oscillator strength  on the amount of the doping
has been calculated in the absence of Coulomb interaction.    
We have found that
$Z^C_{2D}$ for half-filling is  the  same  irrespective of 
the type of CN's and 
the radius of the tube. 
The doping dependence, however, are different 
reflecting their peculiar band structures.  
It is found that $\Delta Z_{3D}^C$  
is independent of the radius of the tube
and the values for both CN's are close to each other. 
The relation may be useful for determining the amount of doping.  
Here we compare the present results and those by the effective mass
theory,
which has been used for calculating  
the optical conductivity
corresponding to the 2D current density for half-filling
in ref.\citen{Ando}. 
By integrating eq.(2.28) in ref.\citen{Ando} in the absence of Coulomb interaction,
one obtains 
$Z_{2D}^{EM} = (2e^2 \gamma)/(2 \pi R) \sum_{\kappa_n}$
for the oscillator strength for half-filling
where $\kappa_n = n/R$ ($n = 0, \pm 1, \cdots)$ for metallic CN's and 
$\gamma = \sqrt{3} t a /2$. 
In principle, 
there are no upper and lower bounds for the transverse momentum $\kappa_n$
in the effective mass scheme. 
However, the number of the allowed transverse momenta
is considered to be proportional to the radius of the tube 
if the cut-off of the order of $a^{-1}$
is introduced. 
Then, $Z_{2D}^{EM}$ is independent 
of the radius, which is qualitatively the same 
as that for the present study. 
However, there remains numerical ambiguity.  

In the present analysis,
we have calculated the doping dependence 
in the absence of Coulomb interaction. 
Though eq.(\ref{eqn:free-sum}) is 
independent of the mutual interaction, 
the oscillator strength of  the tight binding model 
depends on the interaction.\cite{Baeriswyl,Jacobsen,Mila}
Since the long range Coulomb interaction has been considered to play a
crucial roles in CN's,\cite{EG-I,Kane,EG-II,YO,OY,Bockrath} 
it's effect should be investigated. 

The author would like to thank Y. Iwasa for useful comments 
and S. Iwabuchi for stimulating discussion and 
critical reading of the manuscript.
 This work was supported by 
 a Grant-in-Aid 
 for Scientific  Research  from the Ministry of Education, 
Science, Sports and Culture (No.11740196).



\begin{thebibliography}{99}

\bibitem{Iijima}
S. Iijima:
\jo{\NAT}{354}{1991}{56}. 
\bibitem{Hamada}
N. Hamada, S. Sawada and A. Oshiyama:
\jo{\PRL}{68}{1992}{1579}.
\bibitem{Saito-I}
R. Saito, M. Fujita, G. Dresselhaus and M. S. Dresselhaus:
\jo{\PRB}{46}{1992}{1804}.
\bibitem{Saito-II}
R. Saito, M. Fujita, G. Dresselhaus and M. S. Dresselhaus:
\jo{\APL}{60}{1992}{2204}. 
\bibitem{EG-I}  R. Egger and A. O. Gogolin:
\jo{\PRL}{79}{1997}{5082}
\bibitem{Kane}  C. L. Kane, L. Balents, and M. P. A. Fisher: 
\jo{\PRL}{79}{1997}{5086}.
\bibitem{EG-II}  R. Egger and A. O. Gogolin:
\jo{\EPJB}{3}{1998}{281}. 
\bibitem{YO} H. Yoshioka and A. A. Odintsov:
\jo{\PRL}{82}{1999}{374}. 
\bibitem{OY} A. A. Odintsov and H. Yoshioka: 
\jo{\PRB}{59}{1999}{10457}.
\bibitem{Bockrath}  M. Bockrath, D. H. Cobden, J. Lu, 
A. G. Rinzler, R. E. Smalley, L. Balents, and P. L. McEuen: 
 \jo{\NAT}{397}{1999}{598}.
\bibitem{Lee1} R. S. Lee, H. J. Kim, J. E. Fischer, A. Thess
and R. E. Smalley: 
 \jo{\NAT}{388}{1997}{255}.
\bibitem{Rao} A. M. Rao, P. C. Eklund, S. Bandow, A. Thess
and R. E. Smalley: 
 \jo{\NAT}{388}{1997}{257}.
\bibitem{Lee2} R. S. Lee, H. J. Kim, J. E. Fischer, 
J. Lefebvre, M. Radosavljevi${\rm \acute c}$, J. Hone and A. T. Johnson:
 \jo{\PRB}{61}{2000}{4526}.
\bibitem{Claye} A. S. Claye, N. M. Nemes, A. J${\rm \acute a}$nossy, 
and J. E. Fischer:
 \jo{\PRB}{62}{2000}{R4845}.
\bibitem{Wildoer} J. W. G. Wild$\ddot{\rm o}$er, L. C. Venema, 
A. G. Rinzler, R. E. Smalley, and C. Dekker:
\jo{\NAT}{391}{1998}{59}
\bibitem{sum}
See, for example, 
R. A. Bari, D. Adler and R. V. Lange: 
\jo{\PRB}{2}{1970}{2898}, 
I. Sadakata and E. Hanamura:
\jo{\JPSJ}{34}{1973}{882}, 
P. F. Maldague:
\jo{\PRB}{16}{1977}{2437}. 
\bibitem{Ando} 
T. Ando:
 \jo{\JPSJ}{66}{1997}{1066}.
\bibitem{Ichida}
M. Ichida, S. Mizuno, Y. Tani, Y. Saito and A. Nakamura:
 \jo{\JPSJ}{68}{1999}{3131}.
\bibitem{Ando-II}
In ref.\citen{Ando}, it has been pointed out that 
the simple formula, eq.(\ref{eqn:free-sum}), does not hold 
for CN's.  
\bibitem{Baeriswyl}
D. Baeriswyl, J. Carmelo and A. Luther: 
\jo{\PRB}{33}{1986}{7247}. 
\bibitem{Jacobsen}
C. S. Jacobsen: 
\jo{\JPC}{19}{1986}{5643}. 
\bibitem{Mila}
F. Mila: 
\jo{\PRB}{52}{1995}{4788}.

\end{thebibliography}
\end{document}